\documentclass[a4paper]{jpconf}
\newcommand{\p}[1]{(\ref{#1})}
\usepackage{graphicx}
\usepackage{amssymb}
\usepackage{amsmath}

\begin{document}
\title{Relativistic anisotropic stars with the polytropic
equation of state in general relativity}

\author{A A Isayev}

\address{$^1$Kharkov Institute of
Physics and Technology, Academicheskaya Street 1,
 Kharkov, Ukraine}
\ead{isayev@kipt.kharkov.ua}

\begin{abstract}
Spherically symmetric relativistic stars with the polytropic
equation of state, which possess the local pressure anisotropy, are
considered in the context of general relativity. The modified
Lane-Emden equations are derived for the special ansatz for the
anisotropy parameter $\Delta$ in the form of the differential
relation between $\Delta$ and the metric function $\nu$. The
analytical solutions of the obtained equations are found for
incompressible fluid stars. The dynamical stability of
incompressible anisotropic fluid stars
against radial oscillations is studied.
\end{abstract}

\section{Introduction. Basic equations}
\label{I}

It was suggested in Ref.~\cite{ApJ74Bowers} that, despite the
spherically symmetric distribution of matter inside a compact
stellar object, it can be characterized by the local pressure
anisotropy. The analysis of the generalized equations of hydrostatic
equilibrium, allowing for the pressure anisotropy, shows that
anisotropy may have the substantial effect on the maximum
equilibrium mass and gravitational surface redshift.
The pressure anisotropy can be caused by different physical reasons,
such as, e.g., availability of superfluid states with the finite
orbital momentum of Cooper pairs~\cite{PTP93Takatsuka,PRC02IR} or
finite superfluid momentum~\cite{PRC02I,RMP04Casalbuoni}, or
 the presence of strong
magnetic fields inside a
star~\cite{Kh,IY_PRC11,IY_PLB12,JPG13IY,IJMPA14I,PRC15I,PRD15Chu}.
In the present work, we will study spherical 
relativistic anisotropic stars with the polytropic  equation of
state, aiming to obtain the modified Lane-Emden (LE) equations for
the special ansatz for the anisotropy parameter $\Delta$ in the
form of the differential relation between $\Delta$ 
and the metric function $\nu$.  In general case, the obtained
LE equations can be integrated only numerically, but the analytic
solutions can be found for incompressible fluid stars. Then we apply
the Chandrasekhar variational procedure~\cite{ApJ64Chandrasekhar} to
study the dynamical stability
of  incompressible anisotropic fluid stars with respect to 
radial oscillations.

For spherically symmetric stars, the line element is written in the
form (in units with $c=1$):
\begin{align}
ds^2=e^{\nu(r,t)} dt^2-e^{\lambda(r,t)}
dr^2-r^2(d\theta^2+\sin^2\theta d\varphi^2).\label{le}
\end{align}
While the matter distribution inside a star is spherically
symmetric, we allow the existence of the local pressure anisotropy
in its interior with the different values of the radial $p_r$ and
transverse $p_\theta=p_\varphi\equiv p_t$ pressures. The
energy--momentum tensor for a spherical 
anisotropic
star reads
\begin{align}
T_i^k=(\varepsilon+p_t)u_iu^k-p_t\delta_i^k+(p_r-p_t)s_is^k,\label{tik}
\end{align}
where  $\varepsilon$ is the energy density of the system,
$u^i=\frac{dx^i}{ds}$ is the fluid four-velocity, and $s^i$ is the
unit space-like vector with the properties
$s^iu_i=0,\; s^is_i=-1$.
For  the motions in the radial direction, the four-vectors $u^i$ and
$s^i$ have the structure 
$u^i=\biggl(\frac{e^{-\frac{\nu}{2}}}{\sqrt{1-v^2e^{\lambda-\nu}}},\frac{ve^{-\frac{\nu}{2}}}
{\sqrt{1-v^2e^{\lambda-\nu}}},0,0\biggr),\label{ui} 
s^i=\biggl(\frac{ve^{\frac{\lambda}{2}-\nu}}
{\sqrt{1-v^2e^{\lambda-\nu}}},\frac{e^{-\frac{\lambda}{2}}}{\sqrt{1-v^2e^{\lambda-\nu}}},0,0\biggr),
$
where $v=\frac{dr}{dt}$ is the radial velocity. 
In the spherically symmetric case, allowing for the motions in the
radial direction, Einstein equations were written in
Ref.~\cite{LL2}.  The radial component of the equation $T_{i;k}^k=0$
can be represented as
\begin{align}
\dot T_1^0+
{T_1^1}^\prime+\frac{1}{2}T_1^0\bigl(\dot\nu+\dot\lambda\bigr)+\frac{\nu^{\,\prime}}{2}\bigl(T_1^1-T_0^0\bigr)+
\frac{2}{r}\bigl(T_1^1-T_2^2\bigr)=0,\label{ddiv}\end{align}
where $\dot\nu\equiv \frac{\partial\nu}{\partial t}$,
$\nu^{\,\prime}\equiv \frac{\partial\nu}{\partial r}$, etc.
 From Eq.~\p{ddiv},  one can get in the static limit the equation for
the hydrostatic equilibrium in the presence of the pressure
anisotropy in the form
\begin{align}
p_r^{\,\prime}=-\frac{\nu\,'}{2}(\varepsilon+p_r)+\frac{2\Delta}{r},
\quad \Delta\equiv p_t-p_r, \label{prpr}
\end{align}
where $\Delta$ is the anisotropy parameter. For the static
configuration, the interior metric function $\lambda$ reads~\cite{ApJ74Bowers} 
\begin{align}
e^{-\lambda(r)}=1-\frac{2G}{r}m(r), 
\quad r<R, \label{lambda}
\end{align}
where $R$ is the radial coordinate at the surface of a star, and
$m(r)=4\pi\int_0^r
\varepsilon r^2 dr$ is the mass enclosed in the sphere of radius $r$. 
From Einstein equations in the static limit, 
one can find
\begin{align}
\nu\,'=2G\frac{m(r)+4\pi p_r r^3}{r(r-2Gm(r))}.\label{nupr}
\end{align}
Hence, the equation of hydrostatic equilibrium for a spherical 
anisotropic star takes the form:
\begin{align}
p_r^{\,\prime}=-G\frac{(\varepsilon+p_r)(m(r)+4\pi p_r
r^3)}{r(r-2Gm(r))}+\frac{2\Delta}{r}. \label{TOV}
\end{align}
As the boundary condition to Eq.~\p{TOV}, we 
set the radial pressure at the center of a star $p_r(0)=p_{r0}$, and
determine the radial coordinate $R$ at the surface from the
condition $p_r(R)=0$. The total mass then can be calculated as
$M=m(R)$. 
After finding the radial pressure distribution $p_r(r)$ (together
with the mass distribution $m(r)$), the metric functions
$\lambda(r),\nu(r)$ can be determined from Eqs.~\p{lambda},
\p{nupr}. At the boundary $r=R$, the metric functions are matchable
to the exterior vacuum Schwarzschild metric:
\begin{align}
\lambda(R)=-\nu(R)=-\ln\biggl(1-\frac{2GM}{R}\biggr).\label{bc}
\end{align}

In order to solve Eq.~\p{TOV}, 
it is necessary to set the equation of state (EoS) of the system.
Further, as the EoS of the system, we choose the polytropic EoS in
the form~\cite{ApJ65Tooper}:
\begin{align}
p_r=K\varrho^\gamma\equiv K\varrho^{1+\frac{1}{n}},\label{EoS}
\end{align}
where $\varrho$ is the mass density, $K$ is some constant,
 $\gamma$ is the
polytropic exponent, $n$ is the polytropic index. For the
EoS~\p{EoS} 
the energy density $\varepsilon$ is related to the mass density
$\varrho$ and the radial pressure $p_r$ by the equation
$\varepsilon=\varrho+\frac{p_r}{\gamma-1}$~\cite{ApJ65Tooper}. 

It is convenient to introduce the auxiliary dimensionless LE
function $\theta$ according to equations:
\begin{align}
p_r=p_{r0}\theta^{n+1},\quad \varrho=\varrho_0\theta^n,\label{theta}
\end{align}
where $\varrho_0$ is the central mass density. It follows from the
boundary conditions for the radial pressure $p_r$  that
\begin{align}
\theta(0)=1,\quad \theta(R)=0.\label{thetabc}
\end{align}
Then Eq.~\p{prpr} of hydrostatic equilibrium can be rewritten as
\begin{align}
2q_0(n+1)d\theta-\frac{4\Delta
dr}{\varrho_0r\theta^n}+\bigl(1+(n+1)q_0\theta\bigr)d\nu=0,\label{maine}
\end{align}
where $q_0\equiv\frac{p_{r0}}{\varrho_0}$.
 In order to solve the  equation of hydrostatic equilibrium, one
needs to specify the anisotropy parameter $\Delta$. We will suppose
that the presence of the anisotropy parameter $\Delta$ doesn't
change the general form of LE equations for relativistic isotropic
stars~\cite{ApJ65Tooper}, but only can change the coefficients in
these equations. Specifically, we will assume that the parameter
$\Delta$ and the metric function $\nu$ are related by the
differential equation
\begin{align}
-\frac{4\Delta
dr}{\varrho_0r\theta^n}+\bigl(1+(n+1)q_0\theta\bigr)d\nu=(1+\beta
q_0\theta)d\nu, \label{ansatz}
\end{align}
where $\beta$ is some real constant. Substituting Eq.~\p{ansatz}
into Eq.~\p{maine} and integrating it, one can obtain the metric
function $\nu(r)$ in the form
\begin{align}
\nu(r)=\ln\frac{1-\frac{2GM}{R}}{\bigl(1+\beta
q_0\theta\bigr)^\frac{2(n+1)}{\beta}}.\label{nuan}
\end{align}
Let us define the auxiliary function
\begin{align}
u(r)\equiv
\frac{m(r)}{M}=\frac{r}{2GM}\biggl(1-e^{-\lambda(r)}\biggr), \;
u(0)=0, u(R)=1, \label{u}
\end{align}
which 
satisfies the differential equation
$Mu\,'=4\pi\varepsilon r^2$. 
Introducing the dimensionless variable $\xi$ and dimensionless
function $\eta$ by the equations
\begin{align}
r=\alpha\xi, \quad\eta=\frac{M}{4\pi\varrho_0\alpha^3}u,\label{xi}
\end{align}
where $\alpha^2=\frac{q_0(n+1)}{4\pi G\varrho_0}$, equations for the
functions $\theta$ and $\eta$ can be obtained in the form
\begin{align}
\frac{\xi-2(n+1)q_0\eta}{1+\beta
q_0\theta}\xi\frac{d\theta}{d\xi}+\eta+q_0\xi^3\theta^{n+1}=0,\label{thetaprn}\\
\frac{d\eta}{d\xi}=\xi^2\theta^n(1+nq_0\theta).\label{etapr}
\end{align}
Eqs.~\p{thetaprn} and \p{etapr} represent the modified LE equations
for relativistic anisotropic polytropes with the EoS~\p{EoS} and
ansatz~\p{ansatz} for the anisotropy parameter $\Delta$, after
solving which one can find from Eqs.~\p{theta}, \p{xi} the radial
distribution of the radial pressure and mass in the interior of a
spherical anisotropic star. One can see that the obtained LE
equations formally look as in the isotropic case~\cite{ApJ65Tooper},
but with that difference that the impact of the anisotropy parameter
is reflected in the coefficient $\beta$ (substituting the multiplier
$(n+1)$). As follows from Eqs.~\p{thetabc}, \p{xi}, the boundary
conditions for the functions $\theta(\xi)$ and $\eta(\xi)$ read
\begin{align}
\theta(0)&=1,\; \theta(\xi_R)=0,\quad \xi_R\equiv R/\alpha\label{thetabounc}\\
\eta(0)&=0,\;
\eta(\xi_R)=\frac{M}{4\pi\varrho_0\alpha^3}.\label{etabc}
\end{align}

In general case, the LE equations~\p{thetaprn} and \p{etapr} can be
integrated only numerically, but the analytical solutions can be
found for incompressible anisotropic fluid stars, characterized by
the constant density $\varrho=\mathrm{const}$. At $n=0$, solutions
for the functions $\theta(\xi)$  and $\eta(\xi)$, satisfying the
boundary conditions~\p{thetabounc} and \p{etabc}, are given by
\begin{align}
\theta(\xi)=\frac{1}{q_0}\frac{(1+3q_0)\bigl(1-\frac{2q_0}{3}\xi^2\bigr)^\frac{3-\beta}{4}-(1+\beta
q_0)}{3(1+\beta
q_0)-\beta(1+3q_0)\bigl(1-\frac{2q_0}{3}\xi^2\bigr)^\frac{3-\beta}{4}},
\quad \eta(\xi)=\frac{\xi^3}{3}.
\end{align}
The positive root of the $\theta(\xi)$ is
$\xi_R=\sqrt{\frac{3}{2q_0}\biggl[1-\biggl(\frac{1+\beta
q_0}{1+3q_0}\biggr)^{\frac{4}{3-\beta}}\biggr]}.$

\section{Dynamical stability of incompressible anisotropic fluid stars}

Let us consider the stability of spherically symmetric anisotropic
stars with respect to  radial oscillations, assuming that they do
not violate the spherical symmetry. Further we will study the small
radial oscillations and will represent the unknown quantities as
$\varepsilon=\varepsilon^0+\delta\varepsilon,\; p_r=p_{r}^0+\delta
p_r,\; p_t=p_{t}^0+\delta p_t,
\nu=\nu^0+\delta\nu,\; \lambda=\lambda^0+\delta\lambda,$
where $\delta\varepsilon,\delta p_r,\delta
p_t,\delta\nu,\delta\lambda$ are small perturbations with respect to
the corresponding values at the state of hydrostatic equilibrium,
denoted by the upper index "0".  Also, for the small radial
oscillations we will consider that $v\ll1$. Following
Ref.~\cite{ApJ64Chandrasekhar}, it is convenient to introduce a
"Lagrange displacement" $\psi$ by the equation $v=\dot\psi$. We will
assume that all perturbations depend  on time only through the
exponential factor $e^{i\omega t}$, where $\omega$ is the frequency
of radial oscillations. Then, using the linearized form of Einstein
equations~\cite{LL2},  and equation for the conservation of the
total baryon number, the variations $\delta\varepsilon,\delta
p_r,\delta\nu,\delta\lambda$ can be expressed through the Lagrange
displacement $\psi$. Substituting these expressions to the
linearized form of Eq.~\p{ddiv}, in the case of the polytropic EoS
one gets
\begin{gather}
\omega^2
e^{\lambda^0-\nu^0}\bigl(\varepsilon^0+p_{r}^0\bigr)\psi=\frac{2\psi}{r}p_r^{\,0\prime}
-
\frac{2\psi}{r}\Bigl(\gamma\bigl(\nu^{\,0\prime}+\frac{\lambda^{\,0\prime}}{2}+\frac{2}{r}\bigr)+
\frac{2}{r}\Bigr)\bigl(p_t^0-p_r^0\bigr) \label{eigfr}\\+8\pi
Ge^{\lambda^0}p_t^0\bigl(\varepsilon^0+p_r^0\bigr)\psi -
\gamma\frac{d}{dr}\Bigl(\frac{2}{r}\bigl(p_t^0-p_r^0\bigr)\psi\Bigr) 
-\frac{\psi}{\varepsilon^0+p_r^0}\Bigl(p_r^{\,0\prime}-
\frac{2}{r}\bigl(p_t^0-p_r^0\bigr)\Bigr)^2 \nonumber\\ - \gamma
e^{-(\nu^0+\frac{\lambda^0}{2})}\frac{d}{dr}\Bigl(e^\frac{3\nu^0+
\lambda^0}{2}\frac{p_r^0}{r^2}\frac{d}{dr}\bigl(r^2e^{-\frac{\nu^0}{2}}\psi\bigr)\Bigr)
-\frac{2}{r}\Bigl(\gamma
p_r^0\frac{e^\frac{\nu^0}{2}}{r^2}\frac{d}{dr}\bigl(r^2e^{-\frac{\nu^0}{2}}\psi\bigr)+\delta
p_t\Bigr).\nonumber
\end{gather}

Solutions of Eq.~\p{eigfr} for the frequencies of  radial
oscillations should be sought under the boundary conditions
$\psi(r=0)=0,\; \delta p_r(r=R)=0$.
In order to get the variational basis for finding the frequencies
$\omega$, let us multiply both parts of Eq.~\p{eigfr} on
$r^2\psi\exp\bigl(\frac{\nu^0+\lambda^0}{2}\bigr)$ and integrate
over the range of $r$. We will write the corresponding equation
already for incompressible fluid stars  ($n=0$), when the polytropic
exponent $\gamma\rightarrow\infty$. Omitting the upper indexes zero,
one gets
\begin{gather}
\omega^2\int_0^R
e^{\frac{3\lambda-\nu}{2}}\bigl(\varepsilon+p_{r}\bigr)r^2\psi^2\,dr=
\gamma\int_0^Re^{\frac{\lambda+3\nu}{2}}
\frac{p_r}{r^2}\Bigl(\frac{d}{dr}\bigl(r^2e^{-\frac{\nu}{2}}\psi\bigr)\Bigr)^2\,dr
\label{vareq}\\-
\gamma\int_0^Re^{\frac{\lambda+\nu}{2}}r^2\psi\frac{d}{dr}\Bigl(\frac{2}{r}\bigl(p_t-p_r\bigr)\psi\Bigr)\,dr
-2\gamma\int_0^Re^{\frac{\lambda+\nu}{2}}r\psi^2\Bigl(
\nu^{\,\prime}+\frac{\lambda^{\,\prime}}{2}+\frac{2}{r}
\Bigr)\bigl(p_t-p_r\bigr)\nonumber\\
 -2\gamma\int_0^Re^{\frac{\lambda}{2}+\nu}\psi\frac{
 p_r}{r}\frac{d}{dr}\bigl(r^2e^{-\frac{\nu}{2}}\psi\bigr)\,dr.
\nonumber
\end{gather}
In the variational equation~\p{vareq}, the Lagrange displacement
$\psi$ should be chosen such that $\omega^2$ is minimized. If all
frequencies of  radial oscillations are real, a spherical 
anisotropic star is dynamically stable; if some frequency appears to
be imaginary, the configuration is unstable. A~sufficient condition
for the occurrence of the dynamical instability  is
vanishing of the r.-h. s. 
of Eq.~\p{vareq}
for some trial form of the Lagrange displacement $\psi$ satisfying
the boundary conditions.

Let us introduce, following Ref.~\cite{ApJ65Tooper}, the auxiliary
function $\chi=e^{-\frac{\nu}{2}}\psi$. After changing the
integration variable in Eq.~\p{vareq} according to Eq.~\p{xi},
substituting $p_r=q_0\varrho_0\theta$, $\varepsilon=\varrho_0$, and
using the anisotropy parameter $\Delta$ from Eq.~\p{ansatz}  and
expressions~\p{nuan}, \p{u} for the metric functions at $n=0$,
Eq.~\p{vareq}  takes the form
\begin{gather}
\frac{\omega^2}{\omega_0^2}\frac{1}{1-\frac{2GM}{R}}
\int_0^{\xi_R}\frac{(1+q_0\theta)\xi^2\chi^2}{\bigl(1-\frac{2q_0\xi^2}{3}\bigr)^\frac{3}{2}}\frac{d\xi}
{(1+\beta
q_0\theta)^\frac{1}{\beta}}=\gamma\int_0^{\xi_R}\frac{\theta\Bigl(\frac{d}{d\xi}\bigl(\xi^2\chi\bigr)\Bigr)^2}{\xi^2\bigl(1-\frac{2q_0\xi^2}{3}\bigr)^\frac{1}{2}}
\frac{d\xi}{(1+\beta
q_0\theta)^\frac{3}{\beta}}\label{dyneq}\\-\frac{\gamma(1-\beta)}{2}\int_0^{\xi_R}\frac{\xi^2\chi}
{\bigl(1-\frac{2q_0\xi^2}{3}\bigr)^\frac{1}{2}}\frac{d}{d\xi}\Bigl(\frac{\nu'\chi\theta}
{(1+\beta q_0\theta)^\frac{1}{\beta}}\Bigr)\frac{d\xi} {(1+\beta
q_0\theta)^\frac{2}{\beta}} \nonumber
\\-\frac{\gamma(1-\beta)}{2}\int_0^{\xi_R}\frac{\xi^2\chi^2\nu'\theta\bigl(\nu'+\frac{\lambda'}{2}+\frac{2}{\xi}\bigr)}
{\bigl(1-\frac{2q_0\xi^2}{3}\bigr)^\frac{1}{2}}\frac{d\xi} {(1+\beta
q_0\theta)^\frac{3}{\beta}}-2\gamma\int_0^{\xi_R}
\frac{\chi\theta\frac{d}{d\xi}\bigl(\xi^2\chi\bigr)}
{\xi\bigl(1-\frac{2q_0\xi^2}{3}\bigr)^\frac{1}{2}}\frac{d\xi}
{(1+\beta q_0\theta)^\frac{3}{\beta}},
 \nonumber
\end{gather}
where $\omega_0^2=4\pi\varrho_0G$. Let us use the trial functions of
the form $\chi_1=e^{-\frac{\nu}{2}}\xi^2, \chi_2=\xi$.
Then for each given $\beta$ we will try to find such $q_{0c}$ at
which the r.-h. s. 
of Eq.~\p{dyneq} vanishes, and, hence, the
dynamical instability for an incompressible anisotropic fluid star
occurs at $q_0>q_{0c}$.

\begin{table}[tb]
\caption{The critical values of  the parameter $q_{0}$ for the
appearance of the dynamical instability of an incompressible
anisotropic fluid star at different values of the parameter $\beta$.
}
\begin{center}
\begin{tabular}{c|cc}\hline
\multicolumn{1}{c|}{$\beta$}& \multicolumn{2}{c}{$q_{0c}$ evaluated
with the trial function} \\ \cline{2-3} &
\multicolumn{1}{l|}{\hspace{0.8cm}$\chi_1=e^{-\frac{\nu}{2}}\xi^2$\hspace{0.8cm}
} &
\multicolumn{1}{c}{$\chi_2=\xi$} \\ \hline 0.2 & \multicolumn{1}{c|}{2.069} & - \\
{0.4} & \multicolumn{1}{c|}{2.986} & - \\
{0.6} & \multicolumn{1}{c|}{5.189} & - \\
0.8 & \multicolumn{1}{c|}{16.421} & - \\
\hline
\end{tabular}
\end{center}
 \label{table1}
\end{table}

The results of calculations are presented in Table~\ref{table1}. The
most important conclusion is that there are solutions for $q_{0c}$
in the case of the trial function $\chi_1$ at  $\beta<1$, i.e., for
$\Delta=p_t-p_r>0$ (and there are no solutions at $\beta>1$). This
means that the local pressure anisotropy with $p_t>p_r$ can affect
the dynamical stability of spherical 
incompressible
fluid stars, 
unlike to 
incompressible isotropic fluid stars with the polytropic
EoS~\p{EoS}, which are stable against radial
oscillations~\cite{ApJ65Tooper}.

\section*{Acknowledgement} The author would like to thank the
ICPPA--2017 Organizing Committee for invitation and support of his
participation in the conference.

\end{document}